\documentclass{emulateapj}    

\newcommand{\gtapprox}{\raisebox{-0.5ex}{$\,\stackrel{>}{\scriptstyle
\sim}\,$}}
\newcommand{\ltapprox}{\raisebox{-0.5ex}{$\,\stackrel{<}{\scriptstyle
\sim}\,$}}

\usepackage{ulem}

\slugcomment{to appear in ApJ Lett}

\shorttitle{SS433: its circumbinary ring \& system mass}
\shortauthors{Blundell, Bowler \& Schmidtobreick}

\begin{document}

\title{SS433: Observation of the circumbinary disc and extraction of the system mass}

\author{Katherine M.\ Blundell\altaffilmark{1},  Michael G.\
  Bowler\altaffilmark{1} and Linda Schmidtobreick\altaffilmark{2}}

\altaffiltext{1}{University of Oxford, Department of Physics, Keble
  Road, Oxford, OX1 3RH, U.K.}

\altaffiltext{2}{ European Southern Observatory, Vitacura, Alonso de Cordova,
  Santiago, Chile}

\begin{abstract}
The so-called ``stationary" H$\alpha$ line of SS433 is
shown to consist of three components. A broad component is identified
as emitted in that wind from the accretion disc which grows in speed
with elevation above the plane of the disc. There are two narrow
components, one permanently redshifted and the other permanently to
the blue. These are remarkably steady in wavelength and
must be emitted from a circumbinary ring, orbiting the
centre of mass of the system rather than orbiting either the compact
object or its companion: perhaps the inner rim of an excretion disc. The orbiting
speed (approximately 200 km s$^{-1}$) of this  ring material strongly
favours a large mass for the enclosed system (around 40\,$M_\odot$), a large
mass ratio for SS433, a mass for the compact object plus accretion disc of 
$\sim 16$\,$M_\odot$ and hence the identity of the compact object as a rather massive stellar black hole.  
 
\end{abstract}

\section{Introduction}
SS433 is famous for the moving emission lines from its relativistic
precessing jets, but the stationary lines in the spectrum have
enlightened, intrigued and frustrated a generation of astronomers
\citep{Crampton80,Cra81,Fab90,Dodorico91,Gies02}.  Its  binary nature was first indicated by periodic variations in the redshift of stationary H$\beta$
and HeI lines \citep{Crampton80} but the redshifts
were found to be phased incorrectly, relative to eclipses, for these periodic shifts to yield reliably the orbital speed.  Rather, these variations were attributed to H and HeI
lines being formed in a wind or in gas streams within the
system. A blue HeII line may be formed at the
base of the jets, providing a measure of the orbital
speed of the compact object \citep{Fab90,fabrika1997}; see also \citet{Fabrikabook}.

We obtained a sequence of nightly spectra over some 30 days \citep[see][]{BBS07}.  
Our spectra did not extend into the blue but we have studied in detail the brilliant
stationary Balmer H$\alpha$ emission and the daily sampling 
made obvious properties which had
escaped earlier observation.  Over the period JD 2453000 +245.5  to  +274.5
only one observation is missing (+252.5) and in addition SS433 was
quiescent and very well behaved during this period.   After JD +274.5 observations were more
intermittent and at the same time SS433 was becoming restive,
culminating in flaring episodes around JD +294.5.   This paper is
restricted to the period terminating on JD +274.5.

The stationary H$\alpha$ lines possess a profile more complicated than a single gaussian and
so were fitted, spectrum by spectrum, with a number of gaussian components. In
almost every case  three components were required. Where wavelengths or redshifts are quoted in this paper, they refer to the centroids of the fitted gaussians.

\begin{figure}[htbp]
\begin{center}
\centering
   \includegraphics[width=8.5cm]{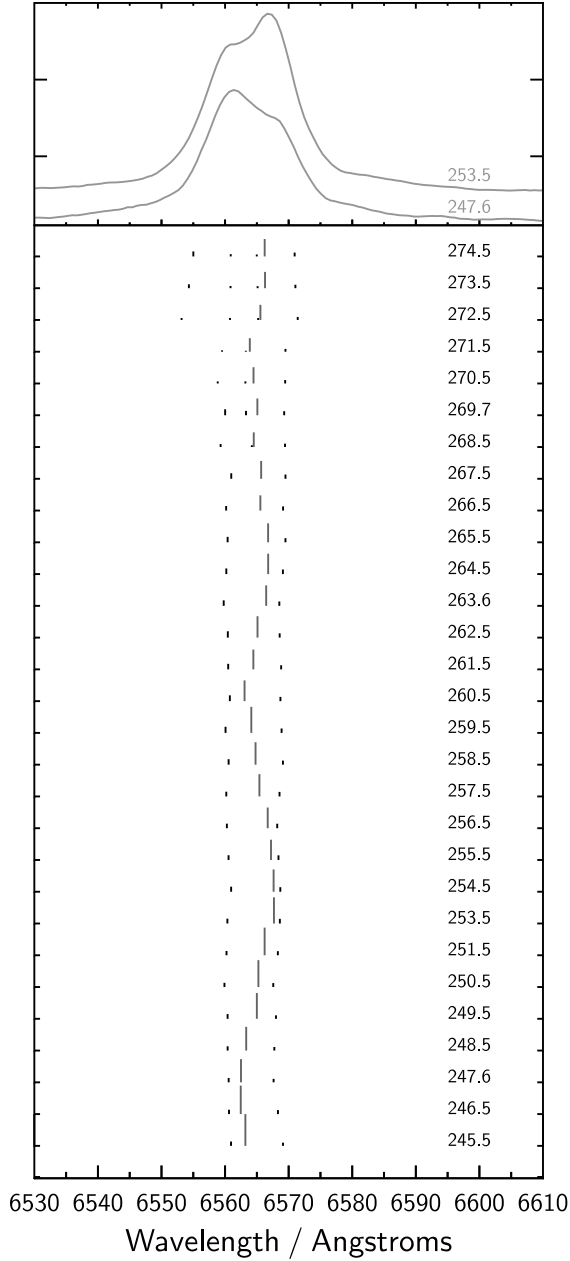} 
\caption{{\it Upper panel:}  examples of two spectra of the stationary Balmer H-$\alpha$ line observed half an orbital period apart.  {\it Lower panel:} wavelengths of components of the gaussians fitted to our sequence of spectra. The Julian Date increases vertically and the heights of the tick marks are the standard deviations of the fitted gaussians. The broad component, in grey, which is identified with the wind from the accretion disc, wanders with 13-day period between the two almost parallel tracks from the circumbinary disc (black). These tracks are separated by almost 400\,km\,s$^{-1}$.  }
\label{fig:newsigma}
\end{center}
\end{figure}

\section{Results}
\subsection{The wind from the accretion disc}
\label{sec:wind}
Fig.\,1 shows the components of the stationary H$\alpha$ line as a
function of time.  The horizontal scale is wavelength, time increases
upwards and the height of each tick is proportional to the standard deviation of the fitted gaussian.  
The nightly observations have created a striking
pattern in this figure: the central component (whose fitted gaussians have standard
deviations as much as 20\,\AA) varies in redshift periodically, 
between two narrow components (having standard deviations of only a few \AA)
which display no significant motion. 
The broad component is very symmetric which implies it
is formed in a fairly transparent wind; Fig.\,2 establishes unambiguously that
this is so.     This figure displays the variation with time of the centre of
the wind component (Fig.\,2a) and of its width, measured by the fitted standard
deviation (Fig.\,2b).   The variation of the central redshift exhibits a
periodicity of approximately 13 days but the width --- a measure of the 
speed of the wind along the line of sight ---  reveals a distinctive pattern with approximately half that periodicity, which matches the nodding period seen in the jets.  Fig.\,2b also shows that the projected wind speed drops as the precessional phase advances (as the jet axes lie more in the plane of the sky and the line-of-sight lies closer to the plane of the disc).  In other data the speed of the wind measured primarily in absorption is slow for a dense equatorial component but otherwise varies as the square of the cosine of the angle between the jets and the line of sight (Fabrika 1997; see also Fabrika 2004, p114). For this reason Fig.\,2c is the cosine of the angle between the jets of SS433 and the line of sight in these same data \citep{BBS07} and the ratio of the width of the wind line to the square of this cosine is shown in Fig.\,2d.  In this ratio the 6.5-day periodicity has disappeared and the ratio is almost flat, apart from a possible 13-day period effect (having maxima at orbital phase 0.55, as for the 13-day jet-speed periodicity \citep{BBS07}). Thus the wind increases in speed with elevation above the plane of the accretion disc.  The dependence of the projected speed of the wind (Fig.\,2b) on both the precessional phase and the nodding of the disc establishes unambiguously that the broad component of the stationary H$\alpha$ line is formed in the wind from the accretion disc.

\begin{figure}[htbp]
\begin{center}
\centering
   \includegraphics[width=8.5cm]{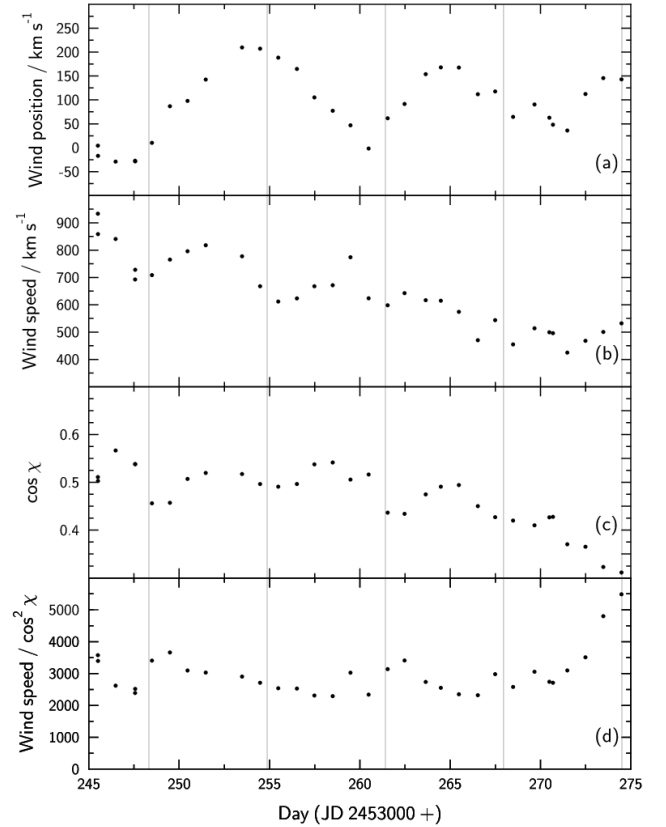} 
\caption{Properties of the wind from the accretion disc. (a) displays the redshift of the wind centroid (grey in Fig.\,1) as a function of time. A 13-day periodicity is apparent but the redshift does not display a clean cosinusoidal variation. In (b) is displayed the standard deviation of the gaussian fitted to the wind; a measure of the speed along the line of sight for a transparent wind. There is a clear 6.5-day periodicity which matches the nodding of the disc. The latter is shown in (c) where $\cos \chi$ (the cosine of the angle the jet axis makes with our line of sight) is plotted, extracted from the nodding of the relativistic jets. Finally (d) displays the result of dividing the wind speed by the square of the cosine of the angle displayed in (c). There may be a residual 13-day oscillation but otherwise the wind speed so corrected is flat (but see text), demonstrating that the wind is rooted in the accretion disc.  The subtle grey lines indicate orbital phases of 0 and 0.5;  phase 0 occurs at Day +254.9.}
\label{fig:wind}
\end{center}
\end{figure}

The amplitude of the 13-day oscillation of the centre of the wind (Fig.\,2a) is $\ltapprox 130$\,km s$^{-1}$ and the maximum redshift occurs just before orbital phase of 0 (eclipse by the
companion) and certainly after orbital phase 0.75, where the compact object is receding most rapidly.   It would be  dangerous to identify this with the speed at which the disc is carried round the system, because the oscillation is not a good sine curve and the amplitude appears to shrink with time.   Further, Fig.\,2d suggests a 13-day periodicity in the wind speed after all the effects of orientation of the disc have been taken out.  The wind from the disc may be interfered with by the larger environment.

\subsection{The circumbinary ring}

The redshifts of the two narrow components of H$\alpha$ are remarkably
constant. The natural interpretation is that we are looking more or
less edge-on to an orbiting ring of glowing material and the narrow
peaks are contributed by the two regions tangential to the line of
sight.  There is no significant 13-day period in the sum of the
redshifts (Fig.\,3a): the source must be {\it circumbinary}.  

This structure cannot be part of the accretion disc. First, the
current estimate of the orbital speed of the compact object is 175\,km
s$^{-1}$ \citep{Fab90,fabrika1997}. Second, the core of the wind
oscillates with a redshift  amplitude of about 130 km s$^{-1}$ and
that wind is rooted in the accretion disc (Sec.\,\ref{sec:wind}).
Finally, in our data after day +274 the accretion disc is unveiled and
observed to orbit the binary centre of mass at approximately 175 km
s$^{-1}$(Blundell, Bowler and Schmidtobreick, in preparation).  

The structure from which the pair of fairly stationary lines emanate
is most unlikely to be a disc orbiting the companion donor star.
However massive the companion may be, it is unlikely to be surrounded
by a stable excretion disc if it fills its Roche lobe, spilling
material onto the compact 
object. Any 13-day periodicity in the mean of the disc redshifts,
phased as for a disc about the companion, is less than 15 km s$^{-1}$.
A donor orbital velocity of 15 km s$^{-1}$ or less about the binary
centre of mass would imply that the mass ratio $ q= m/M$ ($m$ being
the mass of the compact object and $M$ that of the companion) is less
than 0.1 (because $q$ is given by the ratio of the orbital speed of
the companion to that of the compact object). The mass of the compact
object would then be less than 1\,$M_\odot$.  
                                                                                                                                                                                                                                                            \begin{figure}[htbp]
\begin{center}
\centering
   \includegraphics[width=8.5cm]{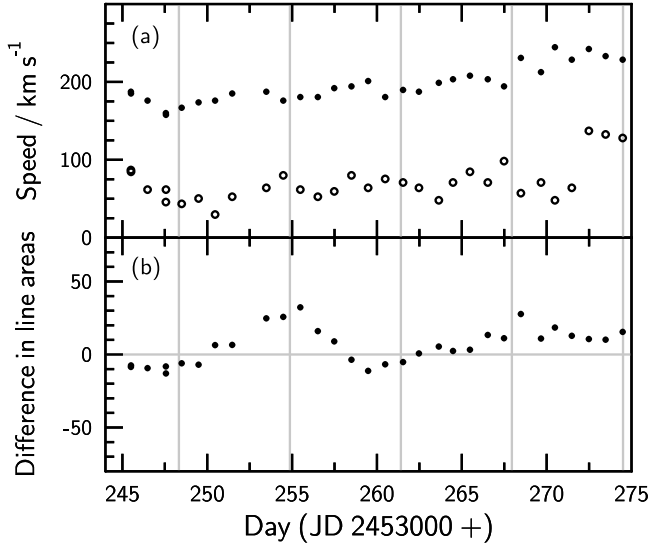} 
\caption{ The upper panel shows the rotational velocity of the disc material, obtained from half the difference of the redshifts of the red and blue components in Fig.\,1 (black). The mean redshift is shown as open circles. In the simple case of both tangent regions radiating with equal intensity both quantities would be constant with time. The lower panel shows the difference between the areas of the gaussians fitted to the red and blue components.  Blue dominates near JD +247 and +260; red in the regions of days +255 and +268.  The subtle grey lines indicate orbital phases of 0 and 0.5;  orbital phase 0 occurs at Day +254.9.} 
\label{fig:statlines}
\end{center}
\end{figure}

Thus the structure radiating H$\alpha$ is almost certain to be circumbinary and the speed with which this material orbits is easily extracted, for the rather constant redshifts and blueshifts shown in Fig.\,1.   Fig.\,3a shows the mean redshift and also half the difference,
which represents the speed at which the material in the glowing ring is
rotating, under the assumption that the centroids of the fitted gaussians correspond to the tangent redshifts.  This appears to increase slightly with time, from 175\,km\,s$^{-1}$ at precessional phase zero to over 200\,km\,s$^{-1}$ at precessional
phase  0.17.   This might be due to projection effects as the true
rotational speed comes closer to the line of sight,  but another explanation is addition of material at smaller radii.  We note that the rotational speed in Fig.\,3a has apparent minima at intervals of about 6.5 days.  The effect is small but not likely to be due to statistical anomalies and we remark it is not an effect of the rotating gravitational quadrupole formed by the compact object and the companion. We estimate that this effect would modulate the rotational velocity with a period of 6.5 days and an amplitude of up to 8 km s$^{-1}$, but that the minimum orbital speeds would be observed at orbital phases of 0.25 and 0.75, rather than approximately 0.5 and 1.  (See also \citet{rudak1981}.) The observed effect is not a real change in local disc velocity, but caused by alternating intensities in the red and blue components, discussed below, and would conceal the tidal effects.

The systemic velocity is about 70\,km\,s$^{-1}$ but there is no indication of
whether this is the real recession velocity of SS433  or
an effect of reddening on the blue edges of the lines. A systemic
velocity of about 70\,km\,s$^{-1}$ is consistent with the HI velocity of the environment (see recent HI data of \citet{lockman2007}), but HeII emission data apparently prefer a systemic velocity close to zero \citep{Fab90,fabrika1997}.   An exotic system such as SS433 may be possessed of a peculiar velocity and the true systemic velocity is hard to establish through optical spectroscopy in a filthy environment. VLBI astrometry reported in \citet{lockman2007} has yielded a radial peculiar velocity relative to the local standard of rest of SS433 of roughly $-17\,{\rm\,km\,s^{-1}}$.

Although these narrow stationary
H$\alpha$ lines stay put, their intensities oscillate with a 13-day period. The oscillations of the two are out of phase, such that when the redshifted line is most intense, the blue is least intense. This phenomenon is clear in the raw spectra \citep{linda2006} 
and quantified in Fig.\,3b, where the
difference in areas of the gaussians fitted to the red and blue lines is plotted as a function
of time. The inference is of ejected hot gas, cooling and fading with time as
the source and ejected gas follow their respective orbits.  This circumbinary structure may feed the large scale feature known as the ruff \citep{Blundell01}. 

Stationary lines from He I and the Paschen H series  are an order of magnitude less intense than the brilliant Balmer H$\alpha$.   They too contain narrow components which are radiated by material in the inner circumbinary ring and the intensities of the red and blue components alternate as the components of H$\alpha$. These lines fade faster than the H$\alpha$ components and do not present as simple a picture as Fig.\,1.   We intend to discuss these at a later date but here simply mention that both He I and the Paschen lines are formed in regions orbiting with speeds of approximately 190 km s$^{-1}$.  The alternating intensities of the Paschen lines have been observed passing through the configuration captured over a few days only by \citet{Filippenko88}. Both the HeI and Paschen lines exhibit much more marked 6.5-day periodicities in the difference velocities than H$\alpha$, because of their stronger alternation in intensities.  (H$\alpha$ and He I can be directly compared in \citet{linda2006}).

\section{Discussion and conclusions}

The oscillating redshift of the overall Balmer H$\alpha$ line \citep{Gies02} is
due to the wind component oscillating in redshift and the red and blue narrow components
alternating in intensity. There may be emissions from gas streams within the binary system itself but the dominant sources of the Balmer H$\alpha$ emission have now been elucidated. 

The glowing ring  contributing the narrow lines is orbiting the system as a
whole, rather than either the compact object or the donor, and the emission fades
with time, the hottest spot matching the rotation of the binary system and
the tail fading, like a katherine wheel. The glowing ring  is most likely to be located at or close to the inner regions of a circumbinary annulus, because the activity requires continual refreshment and is linked to the orbital period of the binary.  The radiating material might either be spilling out piping hot or perhaps disc material excited by X-rays from the accreting system. A further possibility is that the hot spot migrating round the disc is produced where  excreta hit the inner radius of the circumbinary ring.  \citet{Filippenko88} suggested as a possible interpretation of their split Paschen series lines an excretion disc fed by overflow of gas from the $L2$ point in the binary system, with the implication the system is massive, greater
than 40\,$M_\odot$.  Fabrika (1993) remarked such an excretion disc should be visible in H$\alpha$ and the intensities of the two lines should vary with the phase of the orbit; our observations are in excellent agreement with these qualitative predictions.

The masses of the system and the compact object can be extracted from these observations of the circumbinary ring and the orbital speed of the compact object. The radiating material must be orbiting close to the inner stable radius for test particles (or gas)
outside a binary system. This critical distance, $ r_C$, from the binary centre of mass has been
investigated analytically and also using numerical techniques \citep{szebehely1980, rudak1981, Holman99, Artymowicz94}.    For binaries with essentially circular orbits the critical radius is 

\begin{equation}
r_C \approx  f(q) a_B  = f(q)r(1 + q)
\end{equation}
where $a_B$ is the distance between the components of the binary, $q$ is the mass ratio $m/M$ of the components and  $r$ is the radius at which the less massive component orbits the centre of
mass. The value of $f(q)$ is approximately  2;  more precisely, if $f(q) = 2$ at $q=0.1$, it has risen to 2.2 at $q=0.2$ and to  2.3 beyond $q \sim 0.3$ \citep{Holman99}.  For comparison, the ratio of the critical radius to the $L2$ radius is approximately 1.6, roughly independent of the mass ratio $q$ for $q> 0.1$. 

 The speed $v_C$ of Keplerian orbits close to the inner radius of the
 circumbinary ring is given by the relation 

\begin{equation}
v_C = (1 + q) v_X/ \sqrt{f(q)}
\end{equation}
where $v_X$ is the orbital speed of the compact object.

Thus if $f(q)$ is 2 and $v_C \gtapprox v_X$ it follows that the mass
ratio $q > 0.4$.   The mass of the system is given (for an orbital period of 13.08 days) by
  
  \begin{equation}
  M(1+q) = 1.35 f(q)^{3/2} (v_C/100)^3
  \end{equation}
in $M_\odot$, $v_C$ in km s$^{-1}$.

\begin{table}[!h]
\centering
\vspace{0.8cm}
\begin{tabular}{ll|lrrr}
            &    & & \multicolumn{3}{c}{$v_{\rm X}$} \\
            &    & &150\phantom{.0} &    175\phantom{.0} & 200\phantom{.0} \\
\hline
                     &150 & &  5.4                   &     3.7             &   - \\      
$v_{\rm C}$  &175 & & 11\phantom{.0}  &     8.6            &   6.2 \\
                    &200 & & 18.8                   &    15.9            &  12.8 \\
                    &225 & &  30\phantom{.0} &     26.1           &   22.2 \\
\end{tabular}
\caption{\label{tab:two} Mass of compact object as a function of $v_{\rm X}$ and
             $v_{\rm C}$.  Masses are in units of $M_\odot$ and
             speeds in km\,s$^{-1}$.    } 
\end{table}

If the H$\alpha$ radiation from the circumbinary ring 
is coming from
material orbiting at the critical radius or greater, then $v_C$ is greater than or approximately equal to 200 km\,s$^{-1}$. The orbital speed $v_X$ is probably 175 km s$^{-1}$ \citep{Fab90,fabrika1997}.   Thus it  seems  that $q > 0.4$ and  the mass of the
system is $> 16$\,$M_\odot$ and of the compact object $> 6$\,$M_\odot$, including the mass of the accretion disc.   
These are conservative estimates; Table 1, constructed from equations 2 \& 3, 
gives the mass $m$ of the compact object and
its disc as a function of the orbital speed $v_X$ of the compact
object and the speed of the inner circumbinary disc $v_C$.    The smart
money is on the compact object being a rather massive stellar black hole. For $v_X$ = 175 km s$^{-1}$ and $v_C$= 200 km s$^{-1}$ the mass of the compact object is 16\,$M_\odot$ and of the companion 22\,$M_\odot$.


\begin{thebibliography}{}
\bibitem[Artymowicz \& Lubow(1994)]{Artymowicz94} Artymowicz, P., \& 
Lubow, S.H.\ 1994, \apj, 421, 651 

\bibitem[Blundell et al.(2001)]{Blundell01} Blundell, K.~M., 
Mioduszewski, A.J., Muxlow, T.W.B., Podsiadlowski, Ph., \& Rupen, M.P.\ 
2001, \apjl, 562, L79 

\bibitem[Blundell, Bowler \& Schmidtobreick(2007)]{BBS07}
Blundell, K. M., Bowler, M. G. and Schmidtobreick, L. 2007
\aap, 474, 903

\bibitem[Crampton et al.(1980)]{Crampton80} Crampton, D., Cowley, 
A.P., \& Hutchings, J.B.\ 1980, \apjl, 235, L131 

\bibitem[Crampton \& Hutchings(1981)]{Cra81} 
Crampton, D.~\&  Hutchings, J.B.\ 1981, \apj, 251, 604 

\bibitem[D'Odorico et al.(1991)]{Dodorico91} D'Odorico, S., 
Oosterloo, T., Zwitter, T., \& Calvani, M.\ 1991, \nat, 353, 329 

\bibitem[Fabrika(2004)]{Fabrikabook} Fabrika, S.\ 2004, 
Astrophysics and Space Physics Reviews, 12, 1

\bibitem[Fabrika(1997)]{fabrika1997}
Fabrika, S.N., 1997 Astrophysics and Space Sciences, 252, 439

\bibitem[Fabrika(1993)]{fabrika1993}
Fabrika, S.N., 1993, MNRAS, 261, 241

\bibitem[Fabrika \& Bychkova(1990)]{Fab90} 
Fabrika, S.N.~\& Bychkova, L.V.\ 1990, \aap, 240, L5 

\bibitem[Filippenko et al.(1988)]{Filippenko88} Filippenko, A.~V., 
Romani, R.W., Sargent, W.L.W., \& Blandford, R.D.\ 1988, \aj, 96, 242 

\bibitem[Gies et al.(2002)]{Gies02} Gies, D.~R., McSwain, 
M.V., Riddle, R.L., Wang, Z., Wiita, P.J., \& Wingert, D.W.\ 2002, 
\apj, 566, 1069 

\bibitem[Holman \& Wiegert(1999)]{Holman99} 
Holman, M.J., \&  Wiegert, P.A.\ 1999, \aj, 117, 621 

\bibitem[Lockman et al(2007)]{lockman2007}
Lockman, F.J., Blundell, K.M., Goss, W.M. 2007 MNRAS 381 881

\bibitem[Rudak \& Paczynski(1981)]{rudak1981}
Rudak, B. \& Paczynski, B. 1981 Acta Astron., 31, 13

\bibitem[Szebehely(1980)]{szebehely1980}
Szebehely, V. 1980 Celestial Mech., 22 , 7

\bibitem[Schmidtobreick \& Blundell(2006)]{linda2006}
Schmidtobreick, L. \& Blundell, K. 2006 VI Microquasar Workshop: Microquasars and Beyond.
 


\end{thebibliography}
\end{document}